\documentclass[a4paper]{article}

\usepackage{amssymb}
\usepackage{amsmath}
\usepackage{comment}
\usepackage[dvips]{graphicx}
\usepackage{a4wide}
\usepackage{algorithmic}
\usepackage{algorithm} 
\usepackage{epsfig}
\usepackage{url} 
\usepackage{natbib}

\renewcommand{\cite}{\citep}

\newcommand*\oldc{}\let\oldc\c 
\newcommand*\oldH{}\let\oldH\H 

\renewcommand{\c}{\textbf{c}}
\renewcommand{\d}{\textbf{d}}

\renewcommand{\v}{\textbf{v}}

\newcommand{\x}{\textbf{x}}

\newcommand{\z}{\textbf{z}}

\renewcommand{\H}{\textbf{H}}

\usepackage[normalem]{ulem}

\begin{document}

\title{Dissemination of Health information within Social Networks}
\author{Charanpal Dhanjal, Sandrine Blanchemanche, Stephan Clemen\oldc{c}on,\\ Akos Rona-Tas, Fabrice Rossi}
\date{}
\maketitle

\section{Introduction}

State agencies responsible for managing various risks in social life issue advisories to the public to prevent and mitigate various hazards. In this chapter we will investigate, how information about a common food born health hazard, known as Campylobacter, spreads once it was delivered to a random sample of individuals in France. The Campylobacter is most commonly found in chicken meat and causes diarrhoea, abdominal pain and fever. The illness normally lasts a week but in rare cases patients can develop an auto-immune disorder, called Guillain-Barr\'{e} syndrome, that leads to paralysis and can be deadly. Campylobacter together with Salmonella is responsible for more that eighty percent of food born illnesses in France and strikes over 20,000 people each year. People can take simple steps to avoid infection by cleaning their hands, knives, cutting boards and other food items touched by raw chicken meat and by cooking the meat thoroughly.

In this chapter we build two different network models to see how the information about Campylobacter diffuses in society, by mapping onto various network structures the data we gathered with three waves of surveys. In these models the spread of information depends on two sets of factors.  First, each person has a set of individual properties that influences their propensity to transmit the information to or to receive the information from someone they know. Second, each person is connected to others in ways that also affects transmission. There are three aspects of these social ties that matter. As the information travels through existing ties, the quantity of connections should have an influence as people with more ties should have more opportunity to disseminate the information. The quality of ties should matter as well, because certain types of ties may be more conducive to information transmission than others. Finally, the overall structure of the entire network, i.e. how ego-centric networks are linked into a larger whole, should also play a role. Our surveys provide data for the individual characteristics, as well as the quality and quantity of the social ties. In the diffusion model, two different overall network structures, the Erd{\oldH{o}}s and R{\'{e}}nyi random, and the Small World (SW) model, are introduced through modeling assumptions.

The central question of this paper is how individual characteristics and the various aspects of social network influence the spread of information. A key claim of our paper is that information diffusion processes occur in a patterned network of social ties of heterogeneous actors.  Our percolation models show that the characteristics of the recipients of the information matter as much if not more than the characteristics of the sender of the information in deciding whether the information will be transmitted through a particular tie.  We also found that at least for this particular advisory, it is not the perceived need of the recipients for the information that matters but their general interest in the topic. 
         
As for the diffusion of information, we found that the two network structures behave differently in some ways. If the proportion of the population that receives the information initially from the center (our survey) is lower, the random graph model diffuses the information to a larger segment of the population then the SW model, as the advisory travels farther in random networks. However, as the initial exposure increases above a certain level, the two models deliver the advisory by word-of-mouth to very similar proportions of the population. For the SW models we find that as the size of the group that is initially exposed grows, diffusion first increases then decreases and there is an optimal proportion of the population that the message initially has to reach to result in the maximum word-of-mouth diffusion. Since the initial deployment of the information costs money and the subsequent diffusion is costless for the center, finding this optimal size is of practical importance.

Finally, we offer a visual presentation of the transmission process and the way the composition of those receiving the message shifts. Our analysis finds that distribution of particular characteristics of message recipients changes most in the first round and in later rounds it gravitates towards the distribution of the characteristics in the population. 

\section{Theoretical Overview}

\subsection{Diffusion of health information and interpersonal communication}

Major public health campaigns are aimed at modifying individual conduct, by either trying to decrease risky behavior, such as alcohol abuse and smoking or seeking to increase health promoting activities such as exercise and following a healthy diet \cite{compas98sampling}. The approaches to the dissemination of advisories in the field of public health have varied across time, countries and objectives of the campaigns but they are systematically based on specific views of society and on certain assumptions about social processes. 

\subsubsection{Broadcast approach}

The dominant approach in disseminating health advisories has been the broadcast method where the information is released centrally and it is targeted at members of the public individually. The main assumption behind the broadcast approach is that the message travels directly from the center, usually the health agency, to each citizen via various channels of mass communication.  The success of the broadcast approach depends on reaching as many people as possible, and on delivering the message in ways that the desired effects are created in each member of the audience.  The broadcast model is a hub-and-spokes system with the agency in the middle and each unconnected member of the public at the end of one spoke. Communicators using this model, therefore, concentrate on two aspects of the process: reach and stickiness. Reach is the matter of getting the message to the largest possible part of the target audience and involves careful planning of the dissemination of the information. Communicators must decide what media are best suited to their purposes, where they should deploy the information, when and how many times. Stickiness is a matter of creating the proper effect once the information is delivered. Here the main objective is getting people to pay attention to, understand and act on the information in question. To improve individual reception of their messages, broadcasters have spent a lot of effort trying to understand the psychology of message reception and perception in their quest to craft the most effective messages.  

\subsection{The Two-Step Model}

\subsubsection{Mass Communication}

The undisputable strength of the broadcast model is that it provides the maximum control for the center over the message as each recipient obtains the information from the central source. Doubts about the broadcast model, nevertheless, have been raised early on. One of its first critics was the sociologist, Paul F. Lazarsfeld who in a series of studies on political communication in the 1940s found that the direct effect of broadcasted political messages is negligible and that most people acquire their political opinions not from television, radio or newspapers, but from other people he named `opinion leaders.' He argued that political messages produce their effects -- if at all -- in a two step process: the original message is picked up by the opinion leaders who then pass it on to the rest of the community \cite{lazarsfeld48people, katz55personal}. Lazarsfeld's main insight was that most people receive political messages not from the media, from a central point of emission, but by word-of-mouth from others in their community and, therefore understanding the structure of the community is crucial for making communication effective \cite{katz96diffusion, burt99social}. 

\subsubsection{Marketing}

A similar idea surfaced in market research where advertisers encountered the same difficulty in getting their messages across. Researchers found that just as citizens in political discourse, consumers often obtain their information through social ties \cite{katona1955study} and not from the advertisements they are bombarded with. Following this observation, the field of market research has distinguished three types of customers who influence others: the early adopter, the opinion leader and the market maven. Early adopters of new products influence others by buying the product. Their purchases inform other people that the product is available and worth acquiring. Opinion leaders have special expertise about a particular piece of merchandise and dispense it to those interested in the product. Market mavens, on the other hand, are individuals who research and plan their purchases and pay a lot of attention to getting the best deal, and as a result, ``have information about many kinds of products, places to shop, and other facets of markets, and initiate discussions with consumers and respond to requests from consumers for market information'' \cite{feick87market, clark05market}. Marketing people are eager to find early adopters, opinion leaders and market mavens. They focus on the special characteristics of these senders of information. If they can sell them their product (or in the latter two cases sometimes only the idea of the product) they can count on an interpersonal multiplier effect.

\subsubsection{Public Health}

The limitations of the broadcast approach and the research on opinion leaders turned the attention of public health officials to new models of disseminating health information by trying to exploit interpersonal influence. During the 1980s, health promotion programs were increasingly built on community-based intervention that tried to differentiate populations with respect to their health related behavior \cite{shea90review}. This approach drew upon the theoretical perspective of diffusion \cite{rogers03diffusion} and social learning theory \cite{bandura77social} arguing that people acquire new behavior from people in their environment through observational learning.  One well-known example of this approach is the North Karelia Project on smoking. Launched in 1972 in Finland to combat the country's record high mortality of cardiovascular diseases \cite{puska00community}, this project was one of the first major community-based projects for cardiovascular diseases prevention. It was built in partnership with WHO and targeted communities with health information through various channels (television, newspapers, personal health training, seminars, etc.) in an attempt to reduce the number of smokers. Based on the two-step theory of the diffusion of innovations, a network of local opinion leaders was identified in each community often through relevant local organizations. Opinion leaders then were trained to spread the advisory in a credible and effective manner. Unlike broadcasting, community intervention (as its name suggest) aims at a much smaller audience and assumes a simple, two-step connectedness among audience members.

\subsection{Multi-step model}

The concepts of the opinion leader, early adopter and market maven and the strategic actor of the community intervention approach drew attention to the fact that information cannot be thought of as a one step process and that the public is not atomized but linked through social ties. Yet all of them, with the exception of the early adopter, lead only to a two step model. Society or the community is neatly divided between leaders and followers, everyone is either the former or the latter but not both. Studies of early adopters, on the other hand, opened up the possibility of a multi-step diffusion model. As a few, very adventurous early adopters are followed by less adventurous early adopters, and then by mainstream adopters, and finally laggards, the process leads to a chain of diffusion where people in the middle are both leaders and followers.
Empirical studies of diffusion began in the late 1930s and were mostly concerned with the spread of innovation such as the adoption of ham radios \cite{bowers37direction}, progressivist policies \cite{mcvoy40patterns}, hybrid corn \cite{ryan43diffusion}, and the prescription of new drugs by doctors \cite{coleman57diffusion} through imitation. Most studies were interested in measuring the accumulation of adoptions over time. 

In a broadcast model, the curve plotting the cumulative number or proportion of adopters as the broadcast is repeated again and again is close to a logarithmic function truncated at or before the point where no person is left in the population to adopt. The marginal return to broadcasting is highest at the earliest part of the process. The first broadcasts have the highest effect and subsequent ones produce a declining yield.  Once we move away from the broadcast model and allow for imitation, the typical diffusion model becomes an ogive in the shape of an S, as the diffusion process takes off slowly then accelerates when adopters achieve a critical mass or tipping point \cite{schelling78micro, galdwell00tipping}. The curve would slow down as the pool of potential adopters begins to shrink completing the figure. There are several functions that can describe the empirical curve and much of diffusion research consist of correlating curves with the diffusion mechanisms that are thought to have produced them \cite{mahajan1985models}. Until recently, most diffusion models have been macro models predicting only the total number of adopters. As they rarely observe the transmissions, diffusion researchers mostly deduce how they happen from the aggregate outcome by assuming simple transmission rules. The simplest models assume a homogenous population where individuals have different propensities to adopt. Models of the adoption chain start from the assumption that there are personal characteristics that make some people early adopters.  Personal attributes determine at which stage of the diffusion process people enter, i.e. how many adopters they must perceive to make the move. In information diffusion processes, the most curious and cognitively astute people would be the first to find out about an advisory, others would not pick up the information unless they heard it from a friend, yet others would need to hear it from many to understand and believe it etc. Here the spread of information would depend simply on how many people you have of each cognitive type. If you have too few people at the beginning of the chain, they will transmit the information to too few people in the second group, not enough to make anyone care in the third group and the diffusion fizzles out early. If there are more people in the first group, but very few in the second group, the outcome could be similar etc. The thing to notice is that in these models, propensities inherent to each individual that matter \cite{granovetter78threshold, kuran87preference} after that only the aggregate number -- or proportion -- of people who already adopted makes a difference. 
A different approach assumes that all people have the same propensity but they are located differently in a social structure because they are connected differently to others. Whether to adopt or transmit then depends on their relative position -- their spatial \cite{ryan43diffusion, schelling78micro} or social proximity \cite{coleman57diffusion} to those who already adopted.  

In recent years, researchers have moved further away from the broadcast model that assumes that the world can be sharply divided into a source and atomized targets, beyond the two-step model of center, leaders and followers, and past the simplistic, multistep diffusion models and entered a world where all actors are 1) both potential sources and targets, 2) linked in a patterned network of social ties that makes them more or less likely to transmit information and that influences the overall travel of the information in question, and 3) heterogeneous in terms of the properties that makes them more or less likely to send or receive information. In this paper, we will use percolation models that combine graph theory that captures the structural characteristics of networks with dynamic processes.

\subsection{Structural Characteristics of Graphs} 

In order to describe a graph, we distinguish vertices (in our case people) and edges (their ties). To characterize a graph,  it is useful to compute certain properties and we present several useful ones for social networks (for a survey, see \cite{goldenberg01complex}). Perhaps the most well known property of a social network is that they are often `small worlds,' which means that the shortest path length between two people is often small \cite{milgram67sworld}. A formalisation of this measure is the mean geodesic distance for a connected graph, which is the average of the shortest paths between all possible pairs of vertices. Another useful measure for social networks is the clustering coefficient of a graph, which is the chance that a friend of your friend is also your friend.

\subsection{Random Graphs} 

Random graphs are formed by generating a set of edges for a graph in a random fashion. Many random graph models have been extensively studied both theoretically, and in relation to real networks, and later we apply them in our simulated graph processes.  

A classic random graph model is that of Erd{\oldH{o}}s and R{\'{e}}nyi \cite{erdos1959random, erdos1960random}, which was independently discovered by Solomonoff and Rapoport \cite{solomonoff51rand}. In the model, for a graph of size $n$, each pair of vertices has an edge between them with probability $p_e$. One of the properties of the Erd{\oldH{o}}s-R{\'{e}}nyi random graph in the context of social networks is that the clustering coefficient is often quite small. The SW model tends to have a high clustering coefficient and small geodesic distance. It can be constructed by taking a one-dimensional lattice of $n$ vertices in a ring, and joining each vertex to its neighbors $k$ spaces away on the lattice. There are therefore $k n$ edges in this lattice. The edges are re-wired by going through each one in turn and with probability $p_s$ moving one end of the edge to a new vertex chosen uniformly at random. No self edges (edges from a vertex to itself) or double edges (pairs of edges between the same pair of vertices) are created.

An interesting question about random graphs is how they compare to real-world ones. \cite{deker07realistic} finds that real social networks have a low average network distance, a moderate clustering coefficient and an approximate power-law distribution of node degrees. In \cite{newman02randomGraphs} the authors observe that real social networks have highly skewed degree distributions which can vary according to the property being measured. A random graph model is proposed which can be fitted to an arbitrary degree distribution. Such a model is useful since the theoretical computations of clustering and path length over the model often, but not always, bear strong similarities to those found on real data.

\subsection{Modeling Dynamic Processes}

\subsubsection{Disease Spread: Epidemiological Models}  

In epidemiology, there are two deterministic models used most often to study the spread of infectious disease from person to person.  These models are based on a simple mathematical formulation that does not take into account network properties \cite{hethcote00maths}. The SIR model (people are either Susceptible, Infected or Recovered) is an appropriate approximation for diseases that infect a significant part of the population in a short outbreak (such as influenza). This model considers people who recovered from the disease to have acquired permanent immunity. 

In the SIS (Susceptible/Infected/Susceptible) model, people do not acquire permanent immunity and return to the state of susceptibility when they recover from a disease (e.g. tuberculosis). The SIS model is appropriate for endemic diseases which persist in a population for long years. 

Even though the models now being applied to specific diseases are more complex and refined \cite{heathcote89three} ,--  for instance some models use periodic contact rates to take into account the prevalence of many diseases which varies because of seasonal changes in daily contact rates, -- they are quite limited. One of the limits of these basic models is that they make the unrealistic assumption that the population is homogenously or ``fully mixed.'' The homogeneity assumption, that states that everyone is equally susceptible before and infectious after acquiring the sickness, can be relaxed somewhat assuming that the population belongs to a small set of categories (men, women, or adults/children) with heterogenous characteristics. 

These models also assume that the population is equally and randomly connected \cite{watts03six, brauer05kermack}.  Different network structures are addressed by positing multiple levels of mixing, e.g., people may belong to two levels: to a household and to the world, and connected more at one level (with household members) than at another (everyone else) \cite{ball97epidemics}, yet those solutions still miss many dimensions and configurations of social ties.

\subsubsection{Information spread} 

Most empirical studies of information spread investigate diffusion of information through the internet. For instance, information propagation in Weblogs or ``blogs'' is analyzed using a corpus containing 401,012 posts in \cite{gruhl04infoDiffuse}.  One of the dimensions of analysis is the topics of posts. They characterise topics as: ``just spike'' which are inactive, then active, and then inactive again; ``spikey chatter'' which have a significant chatter level and are sensitive to real world event and hence have spikes; and ``mostly chatter'' which have moderate levels of discussion.

The authors also model topic propagation among individuals. An information propagation model is derived based on the Independent Cascade model \cite{goldenberg01complex}. In this extended model, each vertex is a person and each directed edge has a probability of information being copied from one vertex to another in the next time quantum. The model is extended with an additional edge parameter which is the probability that a person reads another persons blog. Edge probabilities for the transmission graph are learnt using an EM-like algorithm \cite{dempster77maxlikelihood}. One of the observations made from the learnt transmission graph is that most people transmit on average to less than one additional person whereas some users transmit to many others, providing a boost to certain topics. An important difference between this work and ours is that individuals can reach many others through their blog posts, and in our case, the number of possible transmissions is limited by the number of regular social contacts.

\cite{kempe03maximisng} consider the problem of selecting the most influential nodes in a social network in the context of information propagation. One application of this work is in the analysis of the ``word-of-mouth'' effect in the promotion of products. The challenge is to discover which individuals should be targeted with information in order to trigger a cascade of further adoption. This problem is NP-Hard, however efficient greedy algorithms are shown to find a solution within sixty-three percent of the optimal for the Independent Cascade and Linear Threshold models. Our diffusion model is essentially a deterministic variant of the information cascade model, and hence this result can also be applied to our work.

\subsubsection{Percolation process}

A percolation process is one in which vertices (sites) or edges (bonds) on a graph are randomly designated either ``occupied'' or ``unoccupied'' and one asks about various properties of the resulting patterns of vertices \cite{newman03graphs}. Percolation theory is mainly developed in physics  but, as \cite{newman03graphs} reminds us, one of the initial motivation of its development in the 1950s was the modeling of the spread of disease and it is still used in epidemiological studies \cite{sander02percolation}. \cite{newmanwatts99scaling} used site percolation on SW graphs as a model of the spread of information or a disease in social networks, and \cite{allard09hetro} deployed a bond percolation model taking into account heterogeneity in the edge occupation probability through a multitype networks approach (see also \cite{reuven00resilience}). Callaway et al. apply a more general approach \cite{callaway00network} that consider the probability of the occupation of a vertex given its degree $k$.

Our paper proposes a model with varied susceptibility to infection and shows under strong heterogeneity in susceptibility there are patches of uninfected but susceptible people. The model uses both bond and site percolation in which vertices either have a specific piece of information or do not. The question in this paper relates to how individuals select among their network of people those interested in the information, and how this information is then diffused to the broader network. 

\section{Data and Method}

There are two approaches to network based research. The first looks at ego networks and takes a sample of individuals and through a series of questions tries to map out their social ties \cite{marsden05recent}. These are then analyzed together with other individual attributes in statistical models assuming independent and random observations. The advantage of ego network research is that it allows for large and representative samples and consequently provides generalizable findings. It also permits the collection of a large amount of information about individuals (egos). This tradition generates data that are strong on vertices, weak and incomplete on edges. The second approach takes entire populations and maps the relationships among all vertices. The relationships or edges are usually complete and they are observed as opposed to reported, but the data about the vertices tend to be limited. The choice of populations -- and therefore the topics -- is opportunistic. 

Our approach belongs to the first line of investigation but with a unique design. We began with an ego network method then we followed the network path through which the information was dispersed collecting data on dyads in some cases from both the sender and the recipient of the information at the two ends of the tie.

\subsection{Data}

To recruit a sample representative for gender, age and socio-economic status of the French population, we first broadly sent an invitation to 24 000 people to answer a questionnaire on "Food Habits and Food Risks". They answered socio-demographic questions which allowed us to select those who fit the quotas. Although the sample is broadly representative for the French population, people in the sample have more frequent connections to the internet than the population as a whole. In the sample, all people use an internet connection whereas 39\% of the French never use one (\cite{jeannine08stats}) and 84,7\% of people in the sample use an internet connection at least once a day while this number is 41\% nationwide.  

The survey was conducted in three waves. The first wave, that took place in December, 2008 and January, 2009 interviewed 6346 individuals, we call Egos. Egos are those who receive the information from us. In our models, Egos can only be senders of the information as they are at the root of the diffusion process. All interviews were conducted through the web using self-administered web surveys. In the first wave, we asked Egos a series of questions about their knowledge of food risks, their food habits, social networks and socio-demographic characteristics. We also exposed them to the information about Campylobacter, which explained the health hazard and provided advice on how to avoid it. We did not tell them that we are interested in information diffusion, but we informed them that we would revisit them for a follow up.

\begin{figure}[htp]
\centering
\includegraphics[bb=0mm 0mm 208mm 296mm, width=99.1mm, height=74.3mm, viewport=3mm 4mm 205mm 292mm]{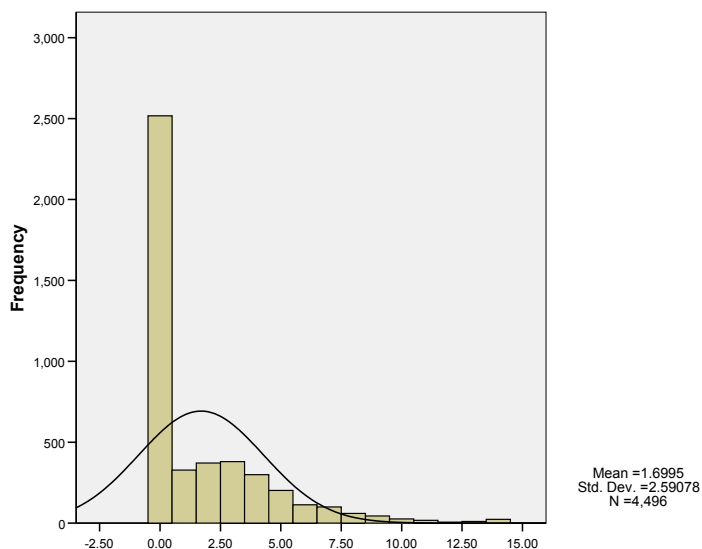}
\caption{The distribution of the number of transmissions reported by Egos}
\label{fig:transmissionPlot}
\end{figure}

We returned to the sample of Egos three weeks later. In the second wave, we were able to interview 4496 of the first wave sample, an attrition of twenty-nine percent. The analysis shows that those who dropped out tended to be somewhat less educated. This time we asked people what they remembered of the information about Campylobacter, how they changed their behavior, and if they told about Campylobacter to anyone else. We asked a series of questions about each person they reported to have talked to about Campylobacter and we requested that by writing in their contact's e-mail address send an email requesting that they fill out a questionnaire for the study. Egos contacted and described 7655 contacts we call Alters. On the average, Egos transmitted the information to 1.7 Alters (Figure \ref{fig:transmissionPlot}). The most common way to convey the information was face-to-face followed by phone conversations. Only less than 6\% of Egos told others about Campylobacter via the internet (Table \ref{tab:transmissionTypes}). None of the socio-demographic variables were strongly correlated to transmission but young people, people working in small companies and those who live with a partner and have children, and those with more than elementary education were a little more likely to transmit the information.  

In the third wave, we interviewed the Alter\index{Alter}s willing to respond to Ego's request. The questionnaire for these Alters was similar to the first and second questionnaires administered to Egos adjusting for differences in context.\footnote{  For instance, we could not ask any questions from Alters that would gauge their knowledge about Campylobacter before they received the information from Ego.  We did ask such questions from the Egos in the first wave before we presented the advisory.} We obtained 451 responses. Only 301 had any recollection of the encounter with Ego. We used only these Alters in our analysis. The responding Alters reported to have contacted yet another 138 people with the information.

\begin{table} 
\centering
\begin{tabular}{|p{1in}|p{0.5in}|} \hline 
Type of contact & Percent \\ \hline 
Face-to-face & 78.95 \\ \hline 
Phone & 15.37 \\ \hline 
Internet (e-mail, Skype etc.) & 5.68 \\ \hline 
Total  & 100 (N=7655)  \\ \hline 
\end{tabular}
\caption{The type of contact through which the information was transmitted }
\label{tab:transmissionTypes}
\end{table} 

There is no generic model for the spread of information independent of the nature of the information to be diffused. In our case, the advisory about the Campylobacter has certain peculiarities. Advisory about this sickness is not knowledge typically sought by its recipients, such as information about chronic diseases or jobs, but rather, it is ``pushed'' by its sender who wants to benefit others (or simply wants to pass the time with small talk). The motivation for transmitting the information is rarely self-interest because the sender typically does not directly benefit from sending the information unless he transmits it to a household member who cooks for him, in which case there is a motivation of self-protection against the disease. 

Furthermore, the transmission is dyadic and does not involve a critical mass. In this respect, it is similar to epidemics where a single contact with an infectious person makes one as sick as multiple contacts with many ill people. Certain types of information do spread better when the sender received it from multiple sources. This is the case when the credibility of the information depends primarily on how widely it is held. Belief in the upward or downward trajectory of the stock market belongs in that category. In our model, the Campylobacter advisory is transmitted dyadically partly because of its content is found to be credible by most people (the average score for both Egos and Alters was 6.1 on a seven-point scale) and there is no need for outside reinforcement.

Moreover, health advisories, like most diseases, are most infectious soon after they are received and as time passes they become less likely to be transmitted . In our data, 91\% of Egos who could recall the timing transmitted the information during the first week and almost half of those on the day they received the information from our survey. Only less than a tenth of the transmissions happened in the second and third weeks. During the time window of our study we observed most of the transmissions that was going to happen and we are unlikely to have missed an outburst of diffusion after we completed our study. 

\section{Computational Simulation} 

There are several limitations of collecting real data, influenced by factors such as cost and privacy. To alleviate these limitations we extract patterns from the survey data and then extrapolate them by simulating a social network. The main aim of this simulation is to measure the average information spread from each Ego on different types of random network. Average information spread is computed using a measure called \emph{average hop length} which is defined as the the total number of transmissions of information divided by the total number of original senders. 

All computational code is implemented using Python and the NumPy, SciPy, Matplotlib, and Mlpy \cite{albanses09mlpy} libraries. The Support Vector Machine (SVM) code is provided using LIBSVM \cite{chang01libsvm}.\footnote{The complete source code for the simulation experiments is available online at http://sourceforge.net/projects/apythongraphlib/.} 

\subsection{Data Preparation} 

In the simulation, vertices represent people and edges are relationships between them. For example, an undirected edge exists between vertices if they know each other as friends, family members, colleagues or acquaintances. Vertices are $\v \in \mathbb{R}^d$, $d=62$, and store values such as the age, gender, education level of each person. Appendix \ref{sec:simulatorFields} shows the complete list of fields used in the simulation. Note that missing values are often replaced with the mode. Furthermore Q7, Q43M1-10, Q47CM1-5 and Q47EM1-5  are categorical variables and hence represented using binary indicator features. As an example, Q7 represents the respondents profession and has 8 categories, hence is represented by 8 binary variables indicating the presence or absence of each category. 

The full details of the data preparation stage are given in Appendix \ref{app:dataCompletion}. Essentially, missing characteristics in the data are completed based on the knowledge aquired about the survey population. The total size of the dataset after the completion step is 86,755 pairs of people (examples), in witch 82,485 were negative (no transmission) and 4,270 were positive (transmission) examples. We denote by $S$ the set of triples composed of pairs of people and an indicator label for transmission occurrence, $S = \{(\v^{(1)}, \v^{(2)}, y): \v \in (E \cup A), y \in \{-1, +1\} \}$, where $y = +1$ indicates information transmission from $\v^{(1)}$ to $\v^{(2)}$.

\subsection{Learning Transmissions}\label{sec:learnTrans}

Given $S$ one must first find a function such that $f'(\v^{(1)}, \v^{(2)}) \approx y$, and we use an SVM \cite{boser92svm} to find this function. To simplify notation let $\x = ({\v^{(1)}}' \; {\v^{(2)}}')' \in \mathbb{R}^{114}$, and the function mapping $\x$ to $y$ be $f$. We will refer to $\x$ as an \emph{example} and $y$ as the corresponding \emph{label}. An SVM finds a hyperplane which separates the set of examples into those  which are positively labelled and those which are negatively labelled. It does this with maximal margin, which often ensures good generalisation to unseen data. A further advantage of SVMs is that they can operate in a kernel defined feature space and hence model non-linear functions without explicit computation of the new feature space. See \cite{shaweTaylor04kernels} for an overview of kernel methods. 

As a first step to learning transmissions, the complete dataset is standardised so that the examples have zero mean and unit standard deviation. They are then randomly sampled into a subset of size 15,000. This subset is used for choosing the SVM penalty parameter C, and the kernel parameters. For this model selection stage, we use a linear kernel with parameter values selected as $C \in \{2^2, 2^3, \ldots, 2^{11}, 2^{12} \}$, and also the  Radial Basis Function (RBF) kernel with $C \in \{2^2, 2^4, \ldots, 2^{10}, 2^{12}\}$,. The RBF kernel is given by $\kappa(\x, \z) = \exp(-\|\x-\z\|^2/2\sigma^2)$ with kernel width $\sigma \in \{2^{-4}, 2^{-3}, \ldots, 2^1\}$ in this case. As the dataset has many more negative examples than positive ones, the penalty on the errors on positive examples is weighted according to $C^{-} = C\gamma$ where $\gamma \in \{2^2, 2^3, 2^4 \}$. 

In order to choose a set of parameters we use $k$-fold Cross Validation (CV) to evaluate prediction error. In this procedure, the dataset is split into $k$ equal sets. One set is kept back for evaluating error, and the remaining are used for training. This is repeated a total of $k$ times with a different test set used each time, and this whole process is repeated for each unique set of SVM parameters. In our case $k=3$. The output at the model selection phase is the set of parameters for the SVM which result in the lowest error. 

Following model selection, one would like to obtain an unbiased estimate of the error obtained on an unseen set of examples. The model selection phase resulted in the selection of the RBF kernel, $C = 2048$, $\sigma=1$ and $\gamma = 32$. We use the set of 71,755 examples disjoint from that used during model selection, and a 5-fold cross validation procedure to obtain an average error with the SVM using this set of parameters. The resulting balanced error is $0.092 \; (0.000)$\footnote{The value in parentheses is the standard deviation of the error.} which compares favourably to the error obtained on predicting no transmissions, which is $0.5$. The error on the positive examples is $0.111 \; (0.000)$ and the error on the negative ones is $0.073\; (0.000)$. With the linear kernel, $C=128.0$ and $\gamma=16$, the best balanced error rate is $0.235 \; (0.000)$ with errors of $0.276 \; (0.000)$ and $0.195 \; (0.000)$ on positive and negative examples. 

\begin{table}[htp] \label{tab:features}
\begin{tabular}{|p{0.2in}|p{0.8in}|p{2in}|p{2in}|p{0.5in}|} \hline 
 & Variable & Meaning & Finding\newline Transmission is more likely if\dots .. & Weight \\ \hline 
\multicolumn{5}{| c |}{Ego's characteristics} \\ \hline 
 & \multicolumn{4}{|p{4.6in}|}{Perception} \\ \hline 
1 & Q17bisA\#X & Risk perception of Campylo (after info) & Ego considers Campylobacter infection  risky  & 76.002 \\ \hline 
 & \multicolumn{4}{|p{4.6in}|}{Knowledge} \\ \hline 
2 & Q16\#X  & Seeking of info (after info) & Ego looked for extra information & -56.268 \\ \hline 
 & \multicolumn{4}{|p{4.6in}|}{Social Network variables} \\ \hline 
3 & Q44BX  & Number of weekly direct contacts with colleagues & Ego~ has fewer direct contacts with colleagues   & -64.349 \\ \hline 
\multicolumn{5}{| c | }{Alter's Characteristics} \\ \hline 
 & \multicolumn{4}{|p{4.6in}|}{Demographic variables} \\ \hline 
4 & Q185\$X & Age & Alter is younger & -51.404 \\ \hline 
5 & Q184\$  & Gender & Alter is female  & -108.127 \\ \hline 
6 & Q186\$  & Education & Alter has less education & -60.097 \\ \hline 
7 & Q187\$\_2 & Profession: enterprise head, artisan, merchant & Alter is not enterprise head, artisan, merchant  & -46.412 \\ \hline 
 & \multicolumn{4}{|p{4.6in}|}{Social Network variables} \\ \hline 
8 & Q178C\$ & Frequency of contact with colleagues & Alter spends more time with colleagues  & -61.636 \\ \hline 
9 & Q178B\$  & Frequency of contact with neighbors & Alter spends more time with neighbors & -57.886 \\ \hline 
10 & Q179M\$\_3 & Group membership parental assoc & Alter is not a member of  parental associations & -51.585 \\ \hline 
11 & Q180C\$X  & Number of weekly direct contacts with family members & Alter contacts many family members  & 46.635 \\ \hline 
12 & Q183EM\$\_2 & Tie with discussion about Food Risk & Alter talks to household member about food risk  & 75.639 \\ \hline 
13 & Q196\$X  & Frequency of internet connection & Alter uses the internet less frequently   & 125.049 \\ \hline 
 & \multicolumn{4}{|p{4.6in}|}{Experience} \\ \hline 
14 & Q4A\$ & Personally cook & Alter does not cook often & -93.724 \\ \hline 
15 & Q24\$X & How often eats chicken (fowl) & Alter eats poultry less often & 48.795 \\ \hline 
 & \multicolumn{4}{|p{4.6in}|}{Knowledge} \\ \hline 
16 & Q1A\$  & General knowledge about food risk & Alter thinks s/he knows more about food risk  & 94.683 \\ \hline 
17 & Q21\$ & Previous knowledge about Campylobacter & Alter had no previous knowledge about Campylobacter & 94.127 \\ \hline 
18 & Q32\$X  & Seeking of info (after info) & Alter did look for more info on Campylobacter & -63.833 \\ \hline 
 & \multicolumn{4}{|p{4.6in}|}{Perception} \\ \hline 
19 & Q33A\$X  & Risk perception of Campylobacter (after info) & Alter considers less risky Campylobacter & -86.062 \\ \hline 
20 & Q20A\$ & Finds info credible &  Alter finds info credible & 49.843 \\ \hline 
\end{tabular}
\caption{Variables in the simulation with the highest weights for predicting transmission. } 
\end{table}

The model weights describe the net influence of the various factors determining whether the information is transmitted from Ego/Sender to Alter/Recipient (Table  \ref{tab:features}). These weights are derived from a model that predicts the pattern of transmission with error. As smaller magnitude weights can be due to prediction error, we concentrate on the large weights and cut out the smaller ones where the relative size of the noise is larger. One of our findings is that in our model, the characteristics of Alters matter more than the characteristics of Egos, as seventeen of the twenty variables with the highest weights belong to Alters. This is surprising because it is the Ego/Sender who decides whether to relay the information. There are two explanations for this finding, one is technical the other is substantive. The technical explanation starts from the recognition that we have considerably fewer fully observed Alters than Egos. The characteristics Alters are predicted and not actually observed. Because we generated the characteristics of recipient and non-recipient Alters differently, and non-transmitting ties always involve on one end a non-recipient Alter and transmitting ties a recipient one, this could have amplified the influence of Alter characteristics on whether or not a tie transmits the advisory. 

The substantive explanation, on the other hand, rests on the assumption that the Egos decide to relay the information on the basis of the characteristics of Alters, although, as we will see not by estimating the recipient's need for the message, which they may not know, but on the basis of their perception of the recipient's general interest in food risk. Another mechanism that can explain the importance of Alters' characteristics is that Recipients elicit the advisory. For instance, women raised to be more attentive to others will be more likely to receive the information by the way they communicate with Egos. Because the technical reasons do play some part in the elevated importance of Recipient characteristics the extent of which we cannot tell, we make our substantive claim with caution.

The factors can be sorted into five categories: demographics, social networks, experience, knowledge and perception. To interpret the effect of social networks on transmission, we have to keep in mind that our unit of analysis is the social tie and not the individual.  The three variables with high weights that describe Ego belong to the perception, knowledge and network categories and our model does not point to any demographic or experiential variables on the Ego side. Two of the three factors that describe transmitting Egos are not surprising. Egos who perceive Campylobacter more risky and those who looked up additional information are more likely to send the information to Alters. The third one, that the frequency of direct contacts with colleagues has a negative effect, will be explained below.

A tie is more likely to transmit the information about Campylobacter if the Alter is female, young, has less formal education and if she is not a self-employed entrepreneur by profession. 

Of the social network characteristics of Alters, spending more time with neighbors, family members and colleagues, increases the chances of a tie to transmit. Interestingly, frequent contacts with colleagues by Ego have the opposite effect. Ties are more likely to transmit if Ego has fewer contacts with, while Alter spends more time with colleagues. This, however, is less of paradox than it seems. 

We observe that this apparent paradox involves two different aspects of collegial ties: quantity and intensity. The second part of this seeming contradiction, that intensity of ties has a positive effect on transmission, is not unexpected. People who spend more time together with colleagues will be more likely to hear about topics unrelated to work.  We find the same relationship for neighbors. 

The second part requires more explanation that involves our unit of analysis: ties and not people. Egos who are well-connected to colleagues in our data set will contribute many collegial ties. This means that if Ego has many collegial ties but only a few will carry the information, most ties with collegially well-connected Egos will not carry the information. When we consider ties, rather than people what matters is not whether the number of transmission rises with the number of ties but whether it rises at a higher rate than ties do.  If the non-transmitting ties rise faster than transmitting ones, the relationship for ties between transmission and the number of ties will be positive for people (the more ties the more transmission) but we will observe a negative relationship for ties (the more ties the less likely a tie will transmit). Collegial ties thus have a diminishing marginal return: the first few colleagues will raise the chances of Ego's telling some of them about the advisory but working with ten as opposed to twenty colleagues makes little difference. 

Why is it then that the number of family contacts increases and not decreases the probability of transmission? The answer is that people have fewer familial than collegial contacts. It seems that the general relationship between number of ties and transmission is such that for the first few ties, not just the number but also the rate of transmission rises. Ego must have a certain number of ties to find one Recipient who is interested, the very first or second tie may not make much of a difference, the third and fourth does. Therefore, what we learn from our model is that higher intensity of collegial and neighborly ties of the recipient will make it more likely that she or he receives the information. The number of ties will yield an increasing return in the beginning but a decreasing one after a certain point. 

We also included measures of formal affiliation to organizations as a measure of social networks.  We learn that transmission is less likely to be targeted at people who are members of parent associations and other formal associations do not seem to make a difference. 

An indirect measure of a type of connectedness is the frequency with which people use the internet. Our survey was conducted through the web, thus it seems strange that people who spend more time on the web are not more but less likely to receive the advisory. Yet this result is consistent with our finding that most of the transmission happens face to face (see Table \ref{tab:transmissionTypes}). 

Experience and practices of Alters seem to indicate that transmission is not driven by the need of the recipient to know about Campylobacter. Those who cook less and eat chicken more often are less likely to receive the information about disease. One possible explanation is that Ego assumes that people who cook a lot and eat lots of chicken are more likely to already be aware of the advisory. What seems to attract the information is if Alter has a general interest in food risk. Ties to recipients who report that they have more general knowledge and then seek additional information on Campylobacter once they received the advisory, i.e., ties to people who are alert and curious about the topic are the most likely to carry the transmission. 

As for Alter's' perceptions, information seems to decrease their fears as the advisory communicates the measures that can be taken to avoid the infection.  This creates a break on the transmission: if perceiving riskiness increases the urge to send the advisory but receiving the information reduces the sense of risk, at the next step, there will be fewer people relaying the information.

\subsection{Modelling Diffusion} 

The SVM model which predicts information transmission is applied to a set of random graphs in this stage. We artificially created a set of vertices use these vertices as a basis of a graph, and select a random subset of the vertices to have information about Campylobacter. Using the SVM model, and various initial setups, we observe how information is diffused within the graph (see Algorithm \ref{alg:simulatorCode} of Appendix \ref{sec:simulatorCode}). 

\begin{figure}[htp]
\begin{center}
        \includegraphics[scale=0.2]{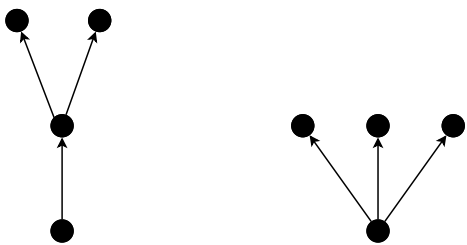}
\end{center}
\caption{An illustration of the difference between $\mu_h$ and $\xi$ measures, in which edges represent information transmissions between vertices. For the graph on the left there is one original sender, two senders and three transmissions and receivers, hence $\mu_h = 3$ and $\xi = 3/2$. For the one on the right there is only one sender and $\mu_h = 3$ and $\xi = 3$.}
\label{fig:hopDiagram}
\end{figure}

At the end of the algorithm, several measurements made during the simulation are output. The quantities $\nu_1, \ldots, \nu_m$ are the proportions of people with information at each iteration. Another quantity of interest is $\xi$ which is the total number of receivers/total number of unique senders. This is a measure of the ``fanout'' of information, i.e. how many vertices receive information directly by each sender on average (see Figure \ref{fig:hopDiagram}).  

The simulation is run using 10,000 vertices, with an SVM trained using 20,000 random examples from the generated transmissions. We use the SW model to generate vertices with neighbors $k \in \{10, 15\}$, re-wiring probability $p_s \in \{0.01, 0.05, 0.1, 0.2\}$ and initial information probability $a \in \{0.1, 0.2, 0.5\}$. Algorithm \ref{alg:simulatorCode} is run five times for each set of parameters and the results are shown in Table  4. The same test is repeated with the Erd{\oldH{o}}s-R{\'{e}}nyi random graph with $p_e \in \{0.001, 0.002, 0.003, 0.004\}$.

\begin{table}
\centering
\begin{tabular}{ l@{ }l  l @{ }l@{ } l @{ }l@{ } l   }
\hline
  $p_e$  & $a$ & $\mu_h$ & $\xi$ & $\nu_1-\nu_0$ & $\nu_2-\nu_1$ & $\nu_3-\nu_2$  \\
  \hline
.001 & .1 & 2.900 (.098) &  1.394 (.015) &  .063(.011) & .031(.015) & .012(.015)\\
.001 & .2 & 2.122 (.081) &  1.141 (.027) &  .090(.011) & .028(.014) & .006(.015)\\
.001 & .5 & 1.460 (.013) &  0.624 (.026) &  .092(.008) & .008(.007) & .001(.006)\\
.002 & .1 & 4.112 (.056) &  1.560 (.016) &  .099(.005) & .056(.010) & .022(.012)\\
.002 & .2 & 2.790 (.065) &  1.148 (.032) &  .130(.010) & .040(.015) & .009(.015)\\
.002 & .5 & 1.829 (.028) &  0.539 (.010) &  .120(.005) & .010(.006) & .001(.006)\\
.003 & .1 & 5.130 (.167) &  1.678 (.036) &  .129(.010) & .073(.018) & .022(.020)\\
.003 & .2 & 3.455 (.102) &  1.210 (.029) &  .159(.009) & .044(.012) & .009(.013)\\
.003 & .5 & 2.225 (.027) &  0.490 (.011) &  .135(.009) & .011(.007) & .001(.006)\\
.004 & .1 & 6.029 (.179) &  1.765 (.034) &  .141(.009) & .078(.016) & .022(.018)\\
.004 & .2 & 4.123 (.033) &  1.264 (.021) &  .178(.007) & .049(.008) & .008(.010)\\
.004 & .5 & 2.721 (.022) &  0.488 (.010) &  .152(.011) & .013(.011) & .001(.010)\\
\hline
\end{tabular}

\caption{Results from the information diffusion simulation using the Erd{\oldH{o}}s-R{\'{e}}nyi model. Standard deviations shown in parentheses. The probability of an edge is $p_e$, the initial information probability is $a$, $\mu_h$ is the average hop distance and $\xi$ is total number of receivers/total number of unique senders. The proportion of vertices with information is recorded as $\nu_0, \ldots, \nu_3$.}
\label{tab:ERSimResults}
\end{table}

\begin{table}
\centering
\begin{tabular}{ l@{ }l @{ } l  l@{ } l @{ }l@{ } l @{ } l  }
\hline
$p_s$ & $k$ & $a$ & $\mu_h$ & $\xi$ & $\nu_1-\nu_0$ & $\nu_2-\nu_1$ & $\nu_3-\nu_2$  \\
\hline
.01 & 10 & .1 & 3.544 (.124) &  1.630 (.065) &  .106(.005) & .040(.006) & .013(.006)\\
.01 & 10 & .2 & 2.677 (.069) &  1.196 (.030) &  .135(.010) & .032(.014) & .006(.015)\\
.01 & 10 & .5 & 1.835 (.031) &  0.528 (.017) &  .124(.016) & .008(.016) & .000(.015)\\
.01 & 15 & .1 & 4.499 (.075) &  1.771 (.050) &  .127(.009) & .060(.013) & .015(.012)\\
.01 & 15 & .2 & 3.363 (.060) &  1.249 (.038) &  .164(.007) & .041(.008) & .006(.009)\\
.01 & 15 & .5 & 2.229 (.041) &  0.493 (.016) &  .136(.006) & .011(.007) & .001(.006)\\
.05 & 10 & .1 & 3.609 (.097) &  1.615 (.039) &  .104(.009) & .043(.011) & .012(.012)\\
.05 & 10 & .2 & 2.676 (.036) &  1.188 (.026) &  .136(.004) & .033(.006) & .007(.006)\\
.05 & 10 & .5 & 1.815 (.034) &  0.526 (.027) &  .119(.005) & .010(.006) & .001(.006)\\
.05 & 15 & .1 & 4.782 (.115) &  1.809 (.031) &  .131(.008) & .063(.012) & .017(.014)\\
.05 & 15 & .2 & 3.343 (.053) &  1.222 (.019) &  .157(.004) & .043(.006) & .007(.009)\\
.05 & 15 & .5 & 2.242 (.031) &  0.497 (.011) &  .137(.003) & .011(.005) & .000(.006)\\
.1 & 10 & .1 & 3.681 (.033) &  1.577 (.021) &  .100(.006) & .044(.008) & .016(.007)\\
.1 & 10 & .2 & 2.713 (.088) &  1.196 (.039) &  .136(.007) & .036(.007) & .007(.008)\\
.1 & 10 & .5 & 1.831 (.016) &  0.528 (.010) &  .121(.009) & .011(.011) & .001(.010)\\
.1 & 15 & .1 & 4.812 (.128) &  1.761 (.031) &  .130(.009) & .061(.015) & .020(.014)\\
.1 & 15 & .2 & 3.373 (.037) &  1.226 (.031) &  .158(.007) & .043(.008) & .007(.008)\\
.1 & 15 & .5 & 2.244 (.057) &  0.495 (.016) &  .139(.005) & .011(.005) & .001(.005)\\
\hline
\end{tabular}
\caption{Results from the information diffusion simulation using the SW model. The re-wiring probability is $p_s$ and $k$ is the initial number of neighbors.}
\label{tab:SWSimResults}
\end{table}

The results with the Erd{\oldH{o}}s-R{\'{e}}nyi graphs are shown in Table 3. In general we observe that the number of new vertices receiving information decreases at each iteration, and one would expect this decrease as the probability that a vertices neighbors also have information increases with $i$. Since  $p_e$ is probability of an edge the mean number of edges is expected to be $n p_e$, and $\mu_h$ should increase with $p_e$. Table \ref{tab:ERSimResults} shows that this is the case, however a doubling in the probability of an edge from 0.1 to 0.2 results in less than double the average hop distance. Note that the more neighbors a vertex has, the more likely that a greater number of those neighbors without information are suitable candidates for transmission (as learnt by the SVM). However, it is clear that as each vertex has more neighbors, the chance of an information-containing vertex coming across a neighbor which also has information increases. In a similar way, an increase in initial information probability corresponds with smaller average hop distances, though a doubling of $a$ results in $\mu_h$ which is greater than half of the original value. 

The values of $\xi$ capture a different aspect of the information diffusion. One would expect a higher value of $a$ to imply more neighbors with information for each vertex and hence a lower $\xi$ value, and in general this is the case. When $a=0.5$, an increase in the value of $p_e$ results in a decrease in $\xi$ possibly since there are more people who are able to send information and fewer who can receive. Notice also that the values of $\nu_{i+1} - \nu_i$  have an interesting trend: for $i=0$ the increase is greatest in most cases when $a=0.2$ compared to when it is either $0.1$ or $0.5$ 

The SW results are given in Table \ref{tab:SWSimResults}. Note that the Erd{\oldH{o}}s-R{\'{e}}nyi graph is similar to a SW model with a re-wiring probability of 1. Hence, a useful comparison is between the Erd{\oldH{o}}s-R{\'{e}}nyi graphs with $p_e = 0.003$ and the SW graphs with $k=15$ and $p_s = 0.1$. The interesting differences in this case are those with the values of $\xi$. The values of $\xi$ are smaller in the Erd{\oldH{o}}s-R{\'{e}}nyi graphs, and since suprisingly the total number of receivers are approximately the same, it implies that there are more unique senders in these graphs. Recall that with high clustering the chance of a friend of a friend being a friend is high and hence fewer senders are able to receive the same number of people as compared with the random connectivity of the Erd{\oldH{o}}s-R{\'{e}}nyi graphs. 

Another interesting finding is that at the lowest level of \textit{a }(\emph{a}=.1)  the Erdos-Renyi graph results in more total number of Alter recipients than the SW network, but as \emph{a} increases, the difference disappears. 
This implies that if the initial broadcasting of the message reaches only a few people, SW networks are less efficient but if the number of Egos gets above a certain proportion of the population there is no difference between the two network types in terms how many Alters the diffusion process delivers. 
 
The overall number of receivers does not vary significantly with changes in $p_s$, however for fixed $p_s$ and $k$ the total number of receivers is highest when $a=0.2$. This latter trend is not generally observed with the Erd{\oldH{o}}s-R{\'{e}}nyi graphs. Clearly, the total number of transmission rises as more edges can accommodate transmissions and falls as the network becomes saturated. The peak is generally before 0.1 with Erd{\oldH{o}}s-R{\'{e}}nyi and approximately 0.2 with SW graphs. 

Several of the other trends present in Table \ref{tab:SWSimResults} are trivial. An increase in $k$ from $10$ to $15$ always results in higher transmission since each vertex has more edges and hence more chance of passing information. The length of information paths is short for high values of $a$, implying that the person who receives information passes it onto many others but those receivers rarely pass it on. Furthermore, when $a=0.5$ the information often only travels along a path of length 2.

In summary this relates to information propagation in the following ways: network structure does not seem to influence the total number of people receiving information. What is important is the number of connections and the probability of having information in the first place. Clearly, not every pair of people will facilitate a transmission and hence if too few people are provided with information, then it may stop before reaching everyone interested in it. Similarly, a saturated network does not permit a lot of transmissions. In the SW model the maximum total number of receivers occurred when 20\% of the population was provided with the information. Since $\nu_{i+1} - \nu_i$ always decreases with $i$, this information is clearly in contrast with ``viral'' information spread in all of the scenarios presented. 

\subsubsection{Transmission Visualisation}\label{section:transmission:visualisation}

The measures recorded in Tables 3 and 4 give a good idea of the information diffusion processes occurring in the generated graphs. We additionally consider the visualisation of graph transmissions. We start with the SW model with $n=10000$, $k=15$, $p_s = 0.1$, and a value of initial information probability of $a=0.1$. In this particular instance, the total number of recipients (including the initial ones) is 2897. Disregarding orientation of information transmission, 2346 persons belong to an unique connected components, while the remaining 551 persons fall into much smaller components (the largest having thirty-five members). However, despite constant progress in graph visualisation \cite{DiBattistaEtAl1999GraphDrawing,HermanEtAl2000Graph}, representing the main connected component in a legible remains impossible. We rely therefore on two simplifying assumptions. 

The first one consists in building a clustering of the nodes of the main connected component \cite{Schaeffer:COSREV2007,FortunatoSurveyGraphs2010}: we find groups of individuals which are more likely to transmit or receive the information inside their group than to members of other groups. Then, rather than displaying the original large graph, we draw a graph of the clusters: each node corresponds to a group of people from the original graph (the surface of the node is proportional to the number of persons in the cluster). The edges between nodes indicate information transmission between members of the corresponding clusters. Concretely, we use maximal modularity clustering \cite{NewmanGirvanModularity2004} with the algorithm described in \cite{Noack2007JGAA,NoackRotta2009MultiLevelModularity} using the implementation provided by Andreas Noack\footnote{Available at \url{http://code.google.com/p/linloglayout/}.}. The visualisation of the clustered graph is done with Fruchterman-Reingold algorithm \cite{FruchtermanReingoldGraph1991} as implemented in the Igraph R package \cite{igraph,RProject}. 

\begin{figure}[htbp]
\centering
\includegraphics[width=0.75\linewidth]{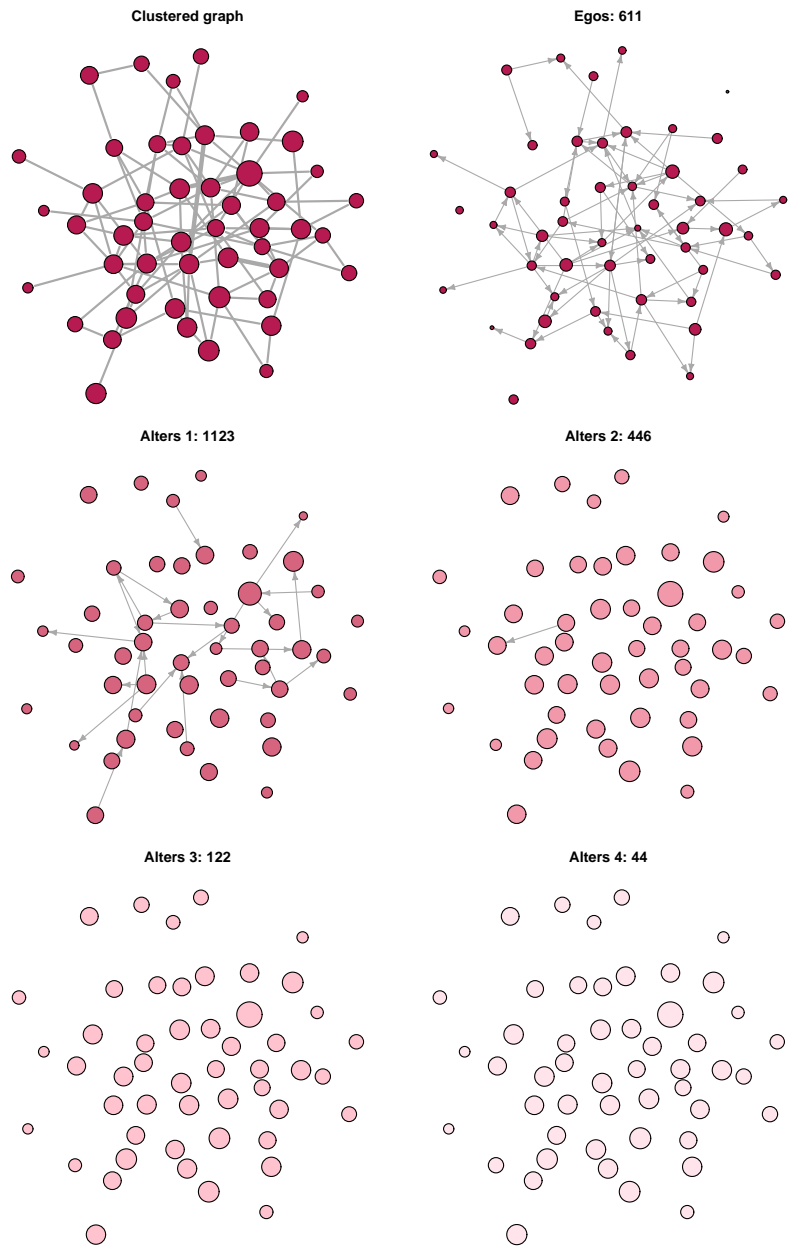}
\caption{A clustered representation of the largest connected component of
  propagation graph used in Section \ref{section:transmission:visualisation}
  (upper left panel) and a clustered representation of information propagation, see main text for details}
\label{fig:clustered:growth}
\end{figure}

The clustering process finds fifty clusters:  less than 3.7 percent of information transmission happens between clusters while the rest takes place inside clusters. This provides a validity index for the clustering: ignoring information propagation outside of a cluster will not introduce major distortions in the analysis. The resulting display is given by Figure \ref{fig:clustered:growth}, upper left panel. The figure shows also information propagation in the graph. Apart from the upper left panel, each graph of this figure shows the number of persons that have received the information at each time step of the propagation (the surface of each node encodes the number of persons). The upper right graph corresponds to the initial receivers (611 Egos), while arrows show information propagation from one cluster to another. The left graph of the second line shows the number of receivers after one step of propagation (i.e. Alters 1), etc. The fact that most of the propagation happens between clusters during the first two steps is easily explained by two factors. Firstly, the initial growth is the largest one and corresponds the largest numbers of transmissions: it should generate the largest part of the external transmissions. Moreover, the clustering algorithm used is based on modularity maximisation. Modularity is a quality criterion for graph clustering that rewards putting connected nodes in the same cluster. However, the reward is inversely proportional to the degree of the nodes. Therefore, the obtained clustering tends to put high degree nodes in different clusters. As the first information transmission is the one in which persons tend to pass knowledge to the largest number of alters, those egos are more likely to be assigned to distinct clusters than the transmitters of the following steps. 

\begin{figure}[htbp]
\centering
\includegraphics[width=0.7\linewidth]{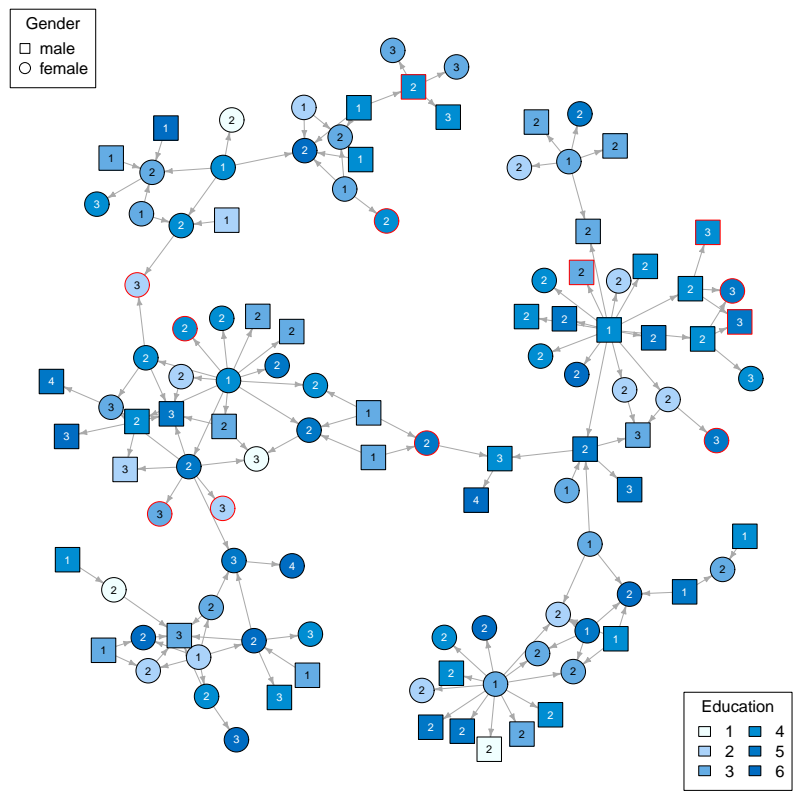}
\caption{Information propagation in the largest cluster, extended as explained in the main text. Education levels are encoded by colours and genders by shape. The iteration number at which the information reached the person is written in the corresponding node. A node with a red border received also the information from a person in a distinct cluster.}
\label{fig:largest:extended:cluster}
\end{figure}

As most of information propagation happens inside clusters, focusing on one cluster provides a good idea of the general transmission. We display information propagation in the largest cluster on Figure \ref{fig:largest:extended:cluster}. To avoid missing information propagation, the cluster was extended in the following way: when information flows \emph{from} a person \emph{in} the cluster under consideration \emph{to} a person in \emph{another} cluster, the recipient is added to the cluster. In the present case, the cluster grows from an initial size of 96 persons to 112 persons. Then, the result presented in Figure \ref{fig:largest:extended:cluster} is exactly the propagation that would have happened even if no other persons apart from those in the cluster had received initially the information. 

\section{Distribution of Individual (Vertex) Characteristics} 

In this test we run a simulation using the SW model with parameters $p_s=0.1$ and $k=15$, and observe how the distribution of various characteristics such as gender, age etc. vary at each iteration.  Transmissions are learnt using a sample of 20,000 examples, and an SVM is trained using the parameters found in Section \ref{sec:learnTrans}. The simulation is run with 20,000 vertices for 3 iterations and repeated a total of 5 times with different random seeds. We observe that there are 1495.2 \emph{\emph{\emph{}}}Egos, and 2689.6, 1245.2, 369.4 new Alters iterations 1, 2 and 3, respectively.

\begin{figure}[htp]
\begin{center}
        \includegraphics[bb=0mm 0mm 208mm 296mm, width=99.1mm, height=53.7mm, viewport=3mm 4mm 205mm 292mm]{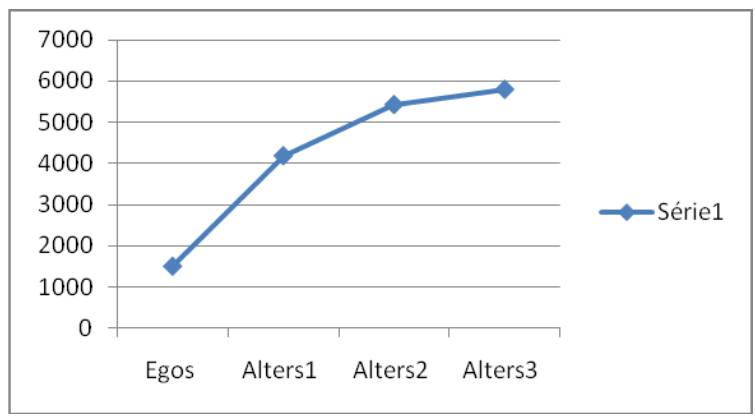}
\end{center}
\caption{Cumulative number in our population who received the advisory in three iterations.}
\end{figure} 

The simulation shows that after the initial jump the number of receivers taper down.  The gender composition changes with each round of iteration. As we have seen earlier women are more likely to be the recipients of the information than men. This explains that there is an overall increase in the proportion of women. This increase mostly levels off after the first round of transmission. This is due not just to the fact that there are increasingly fewer new recipients of the advisory but also that subsequent transmissions are more gender balanced. In the second round the proportion of women actually decreases compared to the first one and levels off in the third, and final round (Table 5).

\begin{table}[ht]
\centering
\begin{tabular}{l l l}
\hline
& Female & Male \\ 
\hline
All & 0.509 & 0.491\\
Egos & 0.535 & 0.465\\
Alters 1 & 0.667 & 0.333\\
Alters 2 & 0.602 & 0.398\\
Alters 3 & 0.598 & 0.402\\
\hline
\end{tabular}
\caption{Distribution of genders in simulated information diffusion.}
\label{tab:genderDistribution}
\end{table}

\begin{figure}[htp]
\begin{center}
        \includegraphics[bb=0mm 0mm 208mm 296mm, width=99.1mm, height=53.7mm, viewport=3mm 4mm 205mm 292mm]{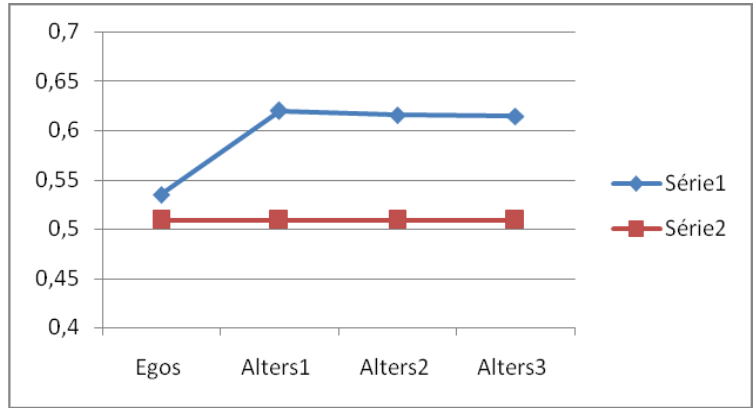}
\end{center}
\caption{Blue: proportion of those who have the information. Red: overall population proportion.}
\end{figure} 

As the information spreads in our simulation we see a similar pattern for the average age of those who receive the advisory (Table 6). (The blue graph in this chart is almost the mirror image of the previous one.) There is an initial drop in age then it changes little. 

\begin{table}[ht]
\centering
\begin{tabular}{l l l l l l l l l l l l l}
\hline
 & 1 & 2 & 3 & 4 & 5 & 6 & 7 & 8 & 9 & 10 & 11 & 12  \\
\hline
All & 0.000 & 0.032 & 0.065 & 0.106 & 0.142 & 0.166 & 0.163 & 0.132 & 0.094 & 0.057 & 0.029 & 0.013\\
Egos & 0.000 & 0.029 & 0.066 & 0.107 & 0.141 & 0.176 & 0.158 & 0.135 & 0.093 & 0.053 & 0.032 & 0.012\\
Alters 1 & 0.000 & 0.041 & 0.072 & 0.120 & 0.154 & 0.169 & 0.162 & 0.121 & 0.083 & 0.047 & 0.021 & 0.011\\
Alters 2 & 0.000 & 0.034 & 0.070 & 0.113 & 0.145 & 0.160 & 0.159 & 0.138 & 0.091 & 0.055 & 0.027 & 0.009\\
Alters 3 & 0.000 & 0.031 & 0.072 & 0.109 & 0.152 & 0.171 & 0.163 & 0.130 & 0.083 & 0.060 & 0.019 & 0.009\\
\hline
\end{tabular}
\caption{Distribution of ages.}
\end{table}

\begin{figure}[htp]
\begin{center}
        \includegraphics[bb=0mm 0mm 208mm 296mm, width=99.1mm, height=53.7mm, viewport=3mm 4mm 205mm 292mm]{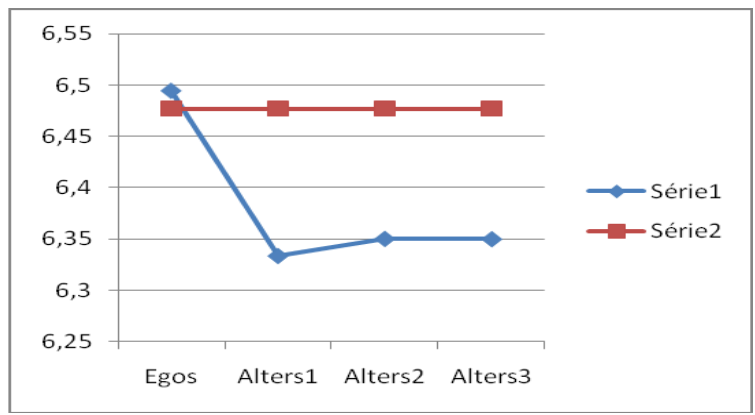}
\end{center}
\caption{Blue: average age of those who have the information. Red: Average age of overall population.}
\end{figure} 

We have similar findings for education and general knowledge of food risk. Overall the pattern is that the decisive change is in the first round and then smaller changes occur for subsequent rounds which together with the ever decreasing number of new recipients results in a stable average. 

\begin{figure}[htp]
\begin{center}
        \includegraphics[bb=0mm 0mm 208mm 296mm, width=99.1mm, height=53.7mm, viewport=3mm 4mm 205mm 292mm]{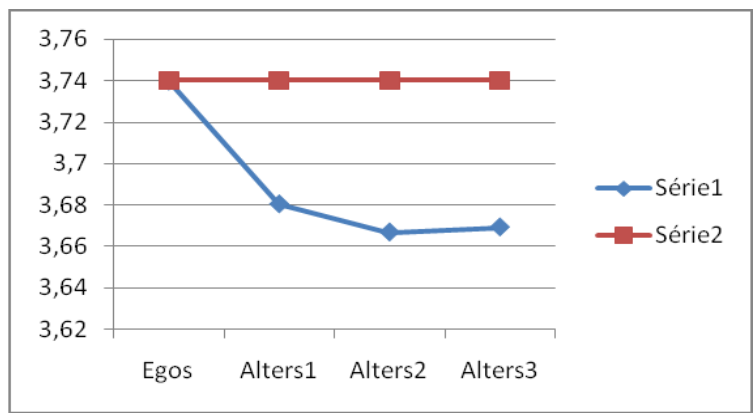}
\end{center}
\caption{Blue: average education of those who have the information. Red: Average education of overall population.}
\end{figure} 

\begin{table}[ht]
\centering
\begin{tabular}{l l l l l l l }
\hline
 & 1 & 2 & 3 & 4 & 5 & 6   \\
\hline
All & 0.030 & 0.123 & 0.265 & 0.313 & 0.199 & 0.069\\
Egos & 0.031 & 0.126 & 0.263 & 0.307 & 0.204 & 0.070\\
Alters 1 & 0.040 & 0.138 & 0.269 & 0.306 & 0.182 & 0.065\\
Alters 2 & 0.039 & 0.140 & 0.277 & 0.309 & 0.175 & 0.060\\
Alters 3 & 0.031 & 0.129 & 0.269 & 0.305 & 0.197 & 0.068\\
\hline
\end{tabular}
\caption{Distribution of education. (1=Without degree/primary/BEPC, 2=CAP/BEP, 3=BAC, 4=BAC+2, 5=More than BAC+2, 6=Doctorate).}
\end{table}

\begin{figure}[htp]
\begin{center}
        \includegraphics[bb=0mm 0mm 208mm 296mm, width=99.1mm, height=53.7mm, viewport=3mm 4mm 205mm 292mm]{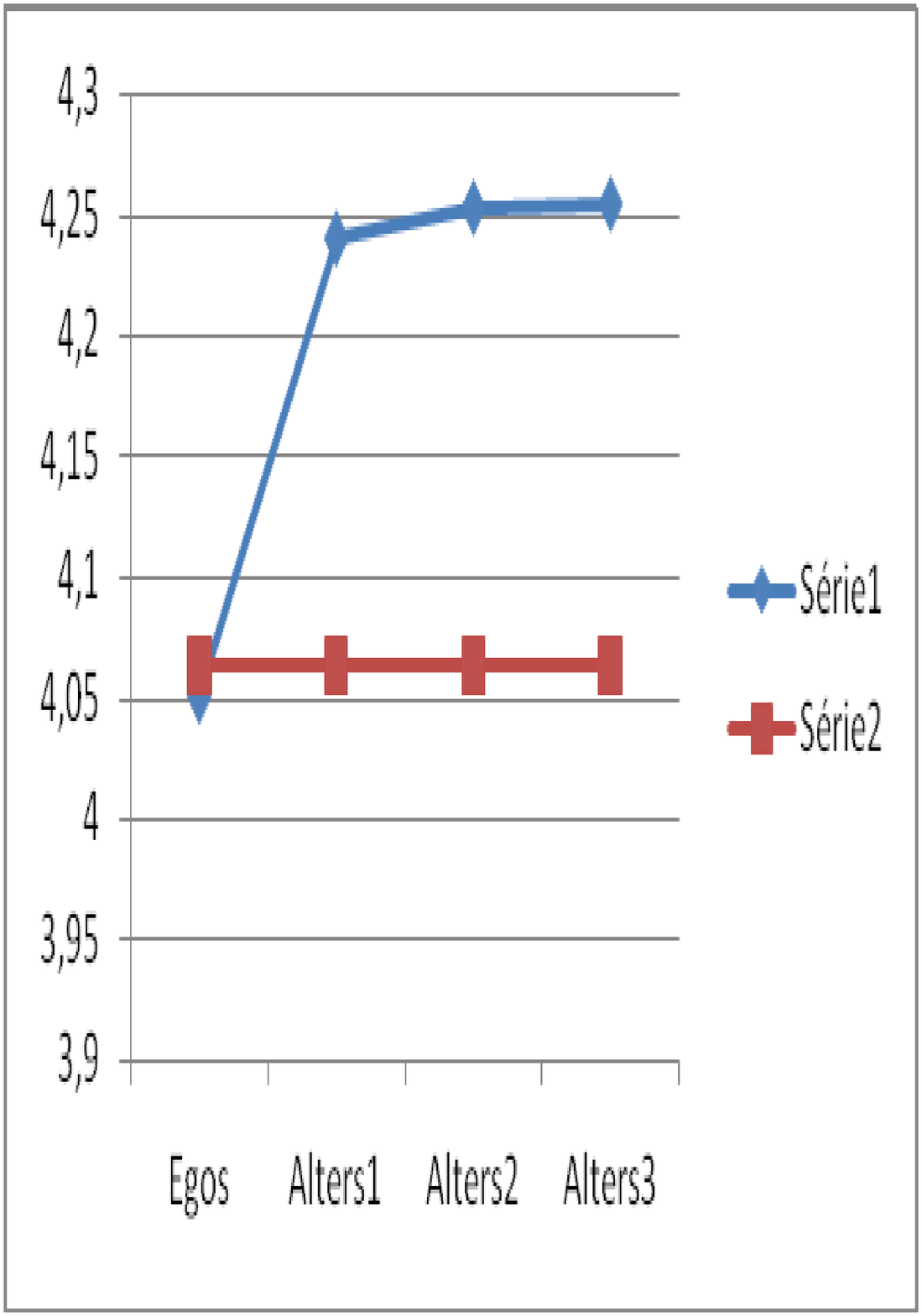}
\end{center}
\caption{Blue: average food risk knowledge of those who have the information. Red: Average food risk knowledge of overall population.}
\end{figure} 

The simulation shows that with the information spreading, the population it reaches is increasingly female, young, less educated and tend to be more knowledgeable about food risks in general, however, much of this shift takes place in the first round of the transmission. This is partly because the characteristics that make people more likely to receive the information are not making them more likely to send it further and it is also partly because of saturation as recipients with those characteristics not having the information in a person`s social circle become more scarce.  

In terms of occupational categories,  the distribution shifts in a way that among the information recipients we find a smaller portion of retirees and people not working at the end of the third round than we had in the beginning (Table 7). The diffusion process reaches the economically active people more successfully than the inactive ones, but, again, we see the same pattern: the first step is the largest one, and then movement is in the opposite direction but in much smaller steps. This pattern is in line with our earlier findings about the role of collegial ties, the type of ties only economically active people possess. Because collegial ties are usually within occupational categories, we can explain the seemingly contradictory findings that net of other factors Alters with less education are more likely to receive the information but the occupational groups with higher average education, such as professionals and employees, will grow faster than others among those who receive the information. It seems that professionals and employees get the advisory not because they are better educated but probably because the nature of the workplace interaction they have compared to workers. Within each group, however, it is the less educated who are more likely to be given the news about the Campylobacter. 

When we rerun the diffusion using the comparable Erd{\oldH{o}}s-R{\'{e}}nyi model, the distribution of the characteristics are very similar and the differences are within the range of random error.

\begin{table}[ht]
\centering
\begin{tabular}{l l l l l l l l l }
\hline
 & 1 & 2 & 3 & 4 & 5 & 6 & 7 & 8  \\
\hline
All & 0.000 & 0.006 & 0.158 & 0.228 & 0.233 & 0.042 & 0.146 & 0.186\\
Egos & 0.000 & 0.005 & 0.160 & 0.217 & 0.245 & 0.040 & 0.157 & 0.174\\
Alters 1 & 0.000 & 0.001 & 0.174 & 0.271 & 0.290 & 0.024 & 0.102 & 0.138\\
Alters 2 & 0.000 & 0.001 & 0.164 & 0.259 & 0.278 & 0.033 & 0.122 & 0.143\\
Alters 3 & 0.000 & 0.002 & 0.175 & 0.238 & 0.273 & 0.033 & 0.128 & 0.151\\
\hline
\end{tabular}
\caption{Distribution of professions.(1=Agricultural workers, 2=Self-employed, 3=Cadres, 4=Professionals, 5=Employees, 6=Workers, 7=Retirees, 8= Other not working.)}
\end{table}

\begin{figure}[htp]
\begin{center}
        \includegraphics[scale=0.7]{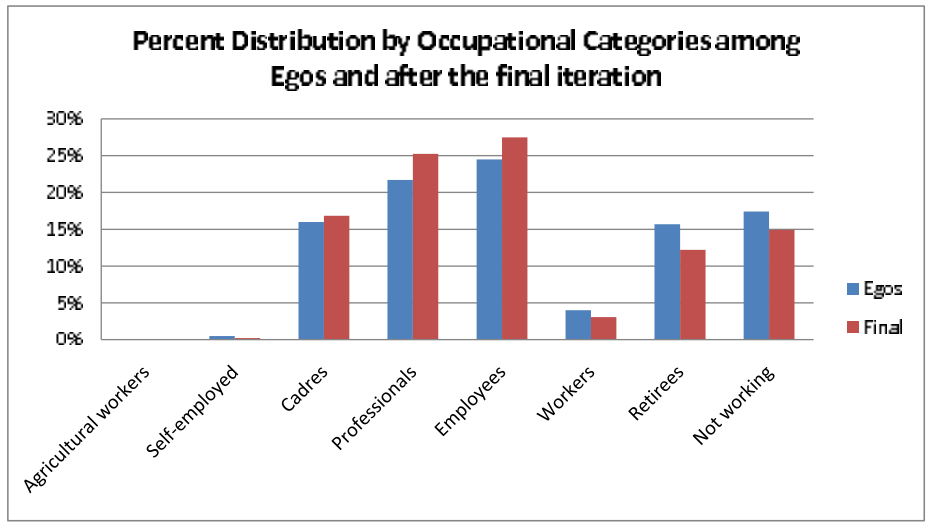}
\end{center}
\caption{Percentage distribution of Occupational categories}
\end{figure}

\begin{figure}[htp]
\begin{center}
        
\end{center}

\end{figure}

\section{Conclusion}

The diffusion of the advisory about the Campylobacter shows that the broadcast model misses an important part of the spread of this type of information; the diffusion of the advisory from its initial audience to others. Unlike the two step process model that, in an attempt to identify opinion leaders, focuses on characteristics make people good senders of the information , we found that the characteristics that make people good recipients are equally, if not more important. Modeling the diffusion showed us that random networks and SW networks produce roughly the same results in terms of the proportion of the population that receives the information but the less clustered random networks achieve the result through longer hop distances and smaller ratio of receivers to unique senders. The composition of those who receive the information changes most in the first transmission and then it levels off not just because there are fewer and fewer recipients making it harder to change the cumulative total but also because new recipients in each round are less different from the overall population. This is partly because the characteristics that make someone more likely to receive the information are not those that make them more likely to pass it on. The mismatch between the two sets of characteristics will dampen transmission even in the absence of network barriers. 

The diffusion of the Campylobacter advisory shows that word-of-mouth dispersion of the information, an aspect of the process the broadcast model ignores, is significant. It also demonstrates that while the first transmission from original Egos to Alters is the biggest part of this word-of-mouth dispersion, the Two-Step model is flawed because there are still substantial -- although smaller -- steps spreading the message it fails to recognize. The Two-Step model also focuses on the qualities of the sender (opinion leader) and our findings show that the recipients' qualities are important. Furthermore, our percolation model was able to build a complex dispersion process which allowed us to account for the heterogeneity of people and of their social networks. Our simulations also showed that not just the local but the global properties of networks matter.  Replicating this study with other advisories will tell us the extent to which our empirical findings can be generalized and be of practical use to state agencies.

\appendix

\section{Data Completion} \label{app:dataCompletion} 

There are $q = 4496$ Egos, denoted by $E = \{\c_1, \ldots, \c_q\}$ with $\c_i \in \mathbb{R}^{62}$ for all $i$. The Egos collectively listed a total of 7655 people with whom they discussed the Campylobacter information. Among them, there are $r = 301$ Alters who responded to the survey for whom we got the set of 62-dimensional vectors, $A = \{\d_1, \ldots, \d_r\}$. The information recorded for the set of other receivers (7204 Alters) is limited to their gender, age, profession and education.\footnote{ An additional 150 Alters failed to remember the information provided to them.ext}  To make a prediction for when information transmission occurs between two people, one needs the complete, immediate, social network for Egos, and the $62$-dimensional vector for Alters. Hence, we completed the data by generating a set of non-receivers and completing the set of receivers (Alters) for each Ego.  

To compute non-receivers, we consider 2 characteristics: one is the number of times Egos have direct contact with friends, colleagues, family members out of the household and acquaintances in a typical week (Q44A-D). The total number of non-receivers is computed using Q44A-D. The other is whether most of the Egos' contacts are homophilic (i.e. they are similar in their characteristics to the Ego, Q46A-D). Homophility can be related to sex, age, education and income. The set of homophiles for the $i$th person $H_i \subseteq E$ is the set of people with identical features as indicated by Q46A-D. For example, if a person $\c_i$ states in Q46A-D that most of their contacts have the same age and gender, then $H_i$ is the set $H_i = \{\c_j : \c_j \in E, \c_{ik} = \c_{jk}, \c_{i\ell} = \c_{j\ell}, i \neq j$\} where $\c_{ik}$ is the $k$th element of $\c_i$, and $k$ and $\ell$ are the indices corresponding to age and gender respectively.  Similarly the set of non-homophiles is $H'_i = E \setminus (H_i \cup \{\c_i\})$. Given the total number of contacts $N_i$ for the $i$th person and a probability $h$ of being a homophile, a set of $h N_i$ vectors is randomly sampled from $H_i$ and $(1-h) N_i$ vectors are sampled from $H'_i$. In our case $h=0.7$. 

As previously stated, the Egos record only the age, gender, profession and education of their Alters, and one needs to complete the data using $A$. For the $i$th person in $E$ and the $j$th Alter recorded, we find a random element of the set of Alter homophiles $G_i = \{\d_j : \d_j \in A, \d_{ik} = \d_{jk}, \d_{i\ell} = \d_{j\ell}, \d_{im} = \d_{jm}\}$ where $\d_{ik}$ is the $k$th element of $\d_i$, and $k, \ell, m$ are the indices corresponding to age, gender and education respectively. 

\section{Information Diffusion Algorithm} \label{sec:simulatorCode}

In Algorithm \ref{alg:simulatorCode} the vertices of the graph are generated using a multivariate normal distribution, with the mean vector and covariance matrix computed using the Egos and Alters from the survey data. Edges in the graph are added according to the Erd{\oldH{o}}s-R{\'{e}}nyi or Small World graph models. At line  \ref{step:random}, $a |V|$ vertices are selected at random and marked as having information. The following inner for loop iterates through all of the edges in the graph, and makes a prediction for whether information is transmitted along that edge using the characteristics of the vertices and the Support Vector Machine (SVM) model learnt at line \ref{step:learnSVM}. Note that the original survey data are not necessarily identically and independently distributed, but we assume so in order to apply the SVM.

\begin{algorithm}
\caption{Pseudo code for information diffusion simulation.}
\label{alg:simulatorCode}
\begin{algorithmic}[1]
\STATE \textbf{Input}: Graph size $n$, iterations $m$, initial proportion with information $a$
\STATE \label{step:learnSVM} Learn SVM model of information transmission
\STATE Create $G = (V, E)$ with $|V| = n$ randomly generated vertices 
\STATE \label{step:random} Randomly select a set $I_0$ of $a |V|$ vertices to mark with information
\FOR{i = 1 to m} 
\FOR{j = 1 to $|E|$} 
\STATE Make information transmission prediction along edge $j$
\ENDFOR 
\STATE $I_i$ is the set of vertices with information, let $\nu_i = |I_i|/n$
\ENDFOR
\STATE Set $\mu_h$ as total number of transmissions/total number of original senders
\STATE Set $\xi$ as total number of receivers/total number of unique senders
\STATE \textbf{Output}: Proportions $\nu_1, \ldots, \nu_m$ of people with information, $\xi$ and $\mu_h$. 
\end{algorithmic}
\end{algorithm}

\section{List of Features Used in Simulation} \label{sec:simulatorFields}

\begin{table}[ht]
\begin{tabular}{ l l}
\hline
  ID & Field Name \\
  \hline
Q4 & Gender \\
Q5 & Age in categories \\
Q48      & Education  \\
Q7 & Profession (8 categories) \\
Q51     & Income \\
Q52     & Single or married \\
Q53     & Number of children \\
Q13     & Size of village/town/city \\
\hline
Q20A &  Personally cook  \\
Q26     & Frequency of eating poultry meat  \\
\hline
Q28A &  General knowledge about food risk \\
Q22     & Personal experience with food risk  \\
Q34 & Previous knowledge about Campylobactor \\
Q16\# & Seeking of info (after info)   \\
Q55     & Frequency of internet connection \\
\hline 
Q17bisA\# & Risk perception of Campylobator \\
\hline
Q37A, Q37B & Finds the information credible, understandable \\
Q37C, Q37D & Finds the information convincing to change behaviour, worrisome \\
\hline 
Q42A-D &        Contact frequency with friends, neighbours, colleagues, family \\ 
Q43M1-10 &      Organisation memberships (1 organisation per field, 10 options)  \\ 
Q44A-D &        Weekly contact with friends, colleagues, family, acquaintances  \\ 
Q45     & Friends closeness \\ 
Q47CM1-5 &      Tie with discussion about Health (6 categories) \\ 
Q47EM1-5 &      Tie with discussion about Food Risk (6 categories) \\ 
Q50     & Number of people at physical place of work     \\ 
Q46A-D & Homophily of contacts (age, gender, education, income) \\ 
\hline
\end{tabular}
\caption{The list of features used for each vertex.}
\label{tab:features:all}
\end{table}

\bibliographystyle{plainnat}

\bibliography{references}

\end{document}